\documentclass[11pt]{article}

\usepackage[normalem]{ulem}
\usepackage{xcolor}
\usepackage[final]{acl}

\usepackage{times}
\usepackage{latexsym}
\usepackage{amssymb}
\usepackage[T1]{fontenc}

\usepackage[utf8]{inputenc}

\usepackage{microtype}
\usepackage{inconsolata}
\usepackage{graphicx}

\usepackage{pifont}
\usepackage{pgfplots}
\pgfplotsset{compat=1.18}
\usepackage{multirow}
\usepackage{caption}
\usepackage{enumitem}
\usepackage{listings}
\usepackage{graphicx}
\usepackage{balance}
\usepackage{hyperref}
\usepackage{tabularray}
\usepackage{cleveref}
\usepackage{threeparttable}
\usepackage{tcolorbox}
\usepackage{xcolor} 
\definecolor{mygreen}{HTML}{2ECC71}
\definecolor{myred}{HTML}{E74C3C}
\definecolor{myblue}{HTML}{3498DB}
\definecolor{mygray}{HTML}{95A5A6}
\newcommand{\cmark}{\textcolor{mygreen}{\ding{51}}}
\newcommand{\xmark}{\textcolor{myred}{\ding{55}}}

\begin{document}

\title{Extending Fill-In-the-Middle with Instructions for Steerable Code Completion}

\author{
  {\bfseries Zhensu Sun$^{1}$ \quad Chengran Yang$^{1,*}$ \quad Chao Peng$^{2}$ \quad Pengfei Gao$^{3}$} \\
  {\bfseries Xiaoning Du$^{4}$ \quad Li Li$^{5}$ \quad David Lo$^{1}$} \\[0.3em]
  {\normalfont $^{1}$Singapore Management University \quad $^{2}$University of Edinburgh \quad $^{3}$Independent Researcher} \\
  {\normalfont $^{4}$Monash University \quad $^{5}$Beihang University} \\[0.3em]
  {\normalfont \texttt{\{zssun,cryang,davidlo\}@smu.edu.sg} \quad \texttt{chao.peng@acm.org}} \\
  {\normalfont \texttt{gaopf1995@gmail.com} \quad \texttt{xiaoning.du@monash.edu} \quad \texttt{lilicoding@ieee.org}} \\[0.3em]
  {\normalfont \small $^{*}$Corresponding author.}
}







\maketitle

\begin{abstract}
Code completion models often fail when the developer's intent is
under-specified in the code context. To mitigate this, developers
frequently use natural language comments to clarify objectives.
However, current code completion models fail to prioritize these
directives effectively since they are merely pre-trained using the
Fill-In-the-Middle (FIM) objective. On the one hand, the natural
language instructions, mixed with the noisy code comments, are just
treated as part of the background context within the prefix. On the
other hand, the pre-training datasets for the FIM objective are mostly
sourced from open-source repositories, which results in a scarcity of
high-intent instruction-to-code pairings that reflect the developers'
workflow in code completion. To bridge this gap, we propose
Instruction-aware Fill-In-the-Middle (IFIM), a fine-tuning method that
extends the FIM structure with a dedicated, structurally separated
instruction section. Our evaluation shows that IFIM substantially
improves adherence to developer intent, while leaving infilling
performance unchanged when no instruction is given. The gains hold on
an in-the-wild benchmark of 100 instructions written by real developers
and across model scales from 1.5B to 7B. IFIM thus offers a backward-compatible upgrade path for existing FIM-based code completion systems at a modest training cost.

\end{abstract}

\section{Introduction}

Modern Integrated Development Environments (IDEs) rely on \emph{code completion} to accelerate development~\cite{wang2023practitioners}: given a partial program, a model predicts the missing sections and displays them to the programmer.
Large Language Models (LLMs) have pushed this feature from keyword-level suggestions to full-context generation, which adapts to different coding styles and languages with remarkable accuracy.
As a result, coding assistants, with LLM-powered code completion as a key feature, have become indispensable tools in the developer community, such as Cursor, Github Copilot, and Trae.

\begin{figure*}[t]
    \centering
    \includegraphics[width=\linewidth]{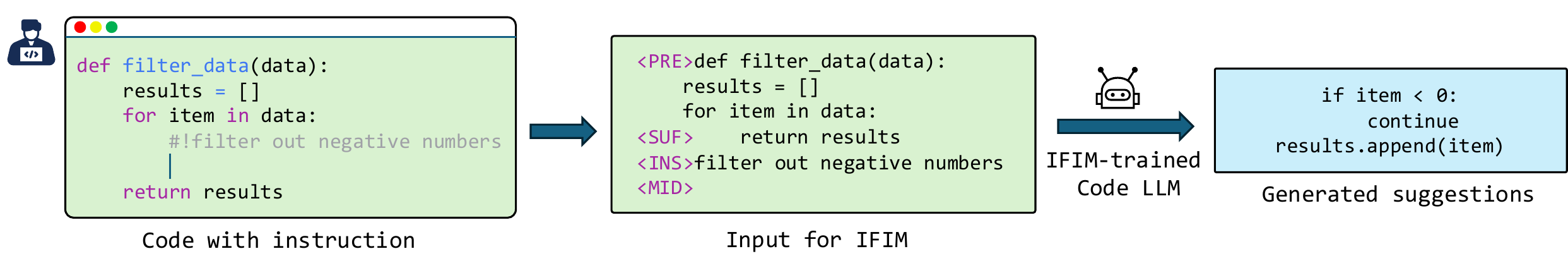}
    \caption{A usage example of IFIM-trained code LLMs. The code context is first processed into IFIM format and then infilled by the model. `\#!' serves as a unique marker (configurable) that distinguishes an IFIM instruction from regular comments, enabling precise parsing.}
    \label{fig:usage}
\end{figure*}

Despite their prevalence, LLM-generated completions are not always helpful. Sun et al.~\cite{sun2025don} revealed that unaccepted completions often stem from \emph{underspecified intent}, i.e., situations where the necessary context is absent from the code.
For such context, it is almost impossible for LLMs, even for human experts, to infer the missing code.
To resolve this ambiguity, developers frequently insert natural language instructions (e.g., inline comments like \texttt{\# filter out negative numbers}) to clarify their goals.
Indeed, studies indicate that programmers spend 11.56\% of their time instructing these LLMs, comparable to the 14.05\% spent on implementing functionality~\cite{mozannar2024reading}. Thus, the ability to accurately follow developer instructions is paramount for the utility of code completion tools.

Due to the unique nature of code completion tasks (i.e., the generated code must integrate correctly back into the original context), the code LLMs adopted for code completion are primarily pretrained models trained using the Fill-In-the-Middle (FIM) objective~\cite{bavarian2022efficient}.
However, FIM is not inherently designed to distinguish between background context and explicit developer intent.
While it allows models to explicitly account for surrounding code, it treats comments containing natural language instructions as ``background noise" within the prefix, indistinguishable from other unrelated comments.
Consequently, models often fail to recognize these instructions as high-priority signals that should override or refine the completion logic.
Furthermore, because most FIM pre-training data is sourced from open-source repositories where explicit ``instruction-to-code'' pairings are scarce, models lack exposure to the specific workflows used by developers to guide code completion models.

To mitigate this, many AI IDEs deploy a secondary LLM specifically to edit code based on instructions.
However, for tasks that a single code completion model could resolve given clear intent, this approach introduces unnecessary architectural complexity, prevents KV-cache reuse, and significantly increases latency.
We argue that enhancing the instruction-following capability of the base code completion model is a more efficient and elegant solution.

A straight-forward solution is instruction-tuning~\cite{wei2021finetuned}: fine-tuning the LLM on (instruction, response) pairs so it learns to follow free-form natural language directives.
Instruction-tuned code LLMs, such as Magicoder~\cite{wei2023magicoder} and Wizardcoder~\cite{luo2023wizardcoder}, can achieve substantial improvements in code generation (i.e., generating a complete function from a natural language description).
However, it backfires for FIM code completion.
LLMs are found to exhibit a degradation in FIM code completion capabilities after the instruction-tuning training step~\cite{pan2024codev,ahmad2025output,wu2024repomastereval}, where an FIM-pretrained base model like CodeGemma-7b~\cite{team2024codegemma} can easily outperform the much larger conversational LLM, GPT-4o, on real-world code completion benchmarks~\cite{pan2024codev}. 
This is because current instruction-tuning methods for code LLMs are designed for generating complete, assistant-style outputs.
It leads to unpredictable behaviors, such as repeating the suffix or introducing unrelated code, in FIM code completion scenarios.
Therefore, though the code LLMs' instruction-following capability can be enhanced, preserving the original FIM capability remains a challenge.

As revealed by OpenAI~\cite{bavarian2022efficient}, the triple-section structure of FIM, i.e., prefix, suffix, and then middle, exhibits a ``FIM-for-free'' property, indicating that fill-in-the-middle can be learned without compromising left-to-right capabilities during pretraining.
Inspired by this property, we investigate whether extending this structure can similarly preserve FIM capabilities while incorporating instructions.
If so, the above-mentioned challenge could be resolved.
Therefore, we propose Instruction-aware Fill-in-the-Middle (IFIM), which extends standard FIM by injecting an instruction section into the input sequence.
IFIM is an off-the-shelf fine-tuning method applicable to existing FIM-trained code models.
As shown in~\Cref{fig:usage}, developers can specify instructions using a distinct marker, allowing the IFIM-trained model to explicitly follow their intent in practice.

To demonstrate the feasibility of IFIM, we constructed a synthesized training dataset using GPT-4.1 (gpt-4.1-2025-04-14) and fine-tuned two base code LLMs, DeepSeek-Coder~\cite{deepseek-ai2024deepseek0coder0v20} and Qwen2.5-Coder~\cite{hui2024qwen25codertechnicalreport}, with the IFIM objective.
The trained models are evaluated on two code completion benchmarks IHumaneval and IRepoMasterEval, derived respectively from Humaneval-infilling~\cite{fried2022incoder} and RepoMasterEval~\cite{wu2024repomastereval} with human-annotated code completion instructions added to their data samples.
Our experiments demonstrate that IFIM achieves remarkable improvements in instruction-following capability while maintaining the LLMs' ability to complete the code without instructions.
Specifically, given the same instructions, our IFIM-trained variants can achieve better performance than the base model, with Pass@1 score averagely improved from 87.8\% to 94.7\% on IHumaneval and from 14.7\% to 20.7\% on IRepoMasterEval.
Crucially, when no instructions are provided, IFIM models either match or slightly exceed their base counterparts' performance.
Beyond clean instructions, IFIM also generalizes robustly to noisy 
developer comments (e.g., typos, verbose phrasing) and transfers 
across six programming languages despite training data being 
predominantly Python. 
Moreover, IFIM introduces negligible inference overhead (0.04\% additional input tokens), preserves suggestion diversity, and maintains clean edit locality, addressing the practical 
deployment concerns of production code completion systems. 
We also conducted a series of ablation experiments to 
analyze the optimal position for inserting the instruction component, the ratio of IFIM processed data, the advantages of using a standalone instruction component and the alignment between 
instructions and generated completions.
These results position IFIM 
as a practical, backward-compatible upgrade path for existing 
LLM-based code completion models.

Our contributions can be summarized as follows:

\begin{itemize}[leftmargin=*]
    \item We show that the three-section FIM structure can be extended with a fourth, structurally separated instruction section without damaging the infilling capability acquired during FIM pre-training. To our knowledge, this feasibility was not previously established, and it opens the door to encoding further completion-time signals as additional sections.
    \item We propose IFIM, a simple yet effective finetuning method that helps FIM-pretrained models better follow natural language instructions without reducing their performance on non-instructed tasks. 
    \item We conduct extensive experiments to show IFIM's dual benefits. Compared with base models, IFIM boosts average Pass@1 scores from 87.8\% to 94.7\% on IHumanEval and 14.7\% to 20.7\% on IRepoMasterEval while maintains performance when no instructions are given.
\end{itemize}

The artifacts are released at
\url{https://github.com/v587su/IFIM_artifacts}.

\section{Related Work}

\subsection{Code LLMs for Code Completion}
Code LLMs have become foundational in automated code completion.
Early models like Codex~\cite{chen2021evaluating}, CodeGen~\cite{nijkamp2022codegen}, and CodeGeex~\cite{zheng2023codegeex} relied on left-to-right (L2R) generation, limiting their ability to handle real-world editing scenarios where code infilling is required.
This gap led to the emergence of FIM training objectives~\cite{bavarian2022efficient,fried2022incoder}, now adopted by state-of-the-art models including CodeGemma~\cite{team2024codegemma}, DeepSeek-Coder-V2~\cite{deepseek-ai2024deepseek0coder0v20}, and Qwen3-Coder~\cite{yang2025qwen3}.
Various studies~\cite{sagtani2025improving, gong2025structure, ding2024horizon} are proposed to enhance FIM.
For example, Sagtani et al.~\cite{sagtani2025improving} introduce curriculum learning and context-aware sampling strategies to enhance FIM-based completion. 
Gong et al.~\cite{gong2025structure} incorporate structural information (e.g., AST-guided masking) into the FIM pretraining process to better align the model with syntactic code editing.
Ding et al.~\cite{ding2024horizon} propose a horizon-length prediction objective to dynamically guide the model’s infilling scope.
Despite these enhancements, both standard and improved FIM methods overlook the instruction-following capability of LLMs, a crucial requirement in real-world code completion scenarios~\cite{mozannar2024reading}.
In this work, we fill this gap by proposing IFIM, a training framework that preserves the infilling capability of FIM while significantly improving the model's ability to follow instructions.

\subsection{Instruction Tuning for Code LLMs}
Instruction tuning, i.e., fine-tuning models on input-output pairs guided by natural language instructions, has proven to be a key technique for improving a model’s ability to follow user intent.
This approach has been widely applied to general-purpose conversational LLMs such as DeepSeek-R1~\cite{guo2025deepseek} and Qwen 3~\cite{yang2025qwen3}.
Instruction tuning has also shown promise in enhancing code LLMs for generation tasks. 
Models like Magicoder~\cite{wei2023magicoder} and WizardCoder~\cite{luo2023wizardcoder}, which are tuned on relatively small instruction datasets, demonstrate strong performance on code generation benchmarks.
However, multiple studies~\cite{ahmad2025output, gong2024evaluation, pan2024codev} reveal that instruction-tuned models remain suboptimal for code completion tasks, when compared to FIM-pretrained base models.
These findings highlight a gap in unifying instruction-following and infilling capabilities.
Inspired by the effectiveness of instruction tuning in natural language domains, we argue that an effective instruction-tuning method tailored for code completion is needed.

\section{Instruction-aware Fill-In-the-Middle}

\subsection{Training Objective}
IFIM further finetunes a FIM-pretrained code LLM, extending its capabilities to understand natural language instructions while preserving its foundational infilling capabilities.
Each training code sample in the training dataset is the program partitioned into three segments, prefix ($P$), middle ($M$), suffix ($S$), with an instruction ($I$) describing the middle segment.
During the finetuning, the model learns to predict the middle segment $M$ conditioned on the combined context of $P$, $S$, and $I$.
This forms a supervised fine-tuning task where the objective is to maximize the log-likelihood of the target sequence:

\[
\max_{\theta} \mathbb{E}{(P,M,S,I) \sim \mathcal{D}} \left[ \log P_{\theta}(M | P, S, I) \right]
\]
where $\mathcal{D}$ is the training dataset, and the input is formatted according to a specific ordering of the components.

\subsection{Training Sample Construction}
Based on the training objective, we formalize the construction of input-output pairs for instruction tuning.
The ground-truth output is the middle segment $M$.
The input needs to be formatted following a specific mode, which specifies how the rest three components are represented.
We term it as IFIM mode.

The base code LLMs to be tuned have their own FIM modes, where prefix, suffix, and middle are concatenated following a specific order with special tokens prepended to the beginning of each section.
For example, a PSM mode indicates the order prefix, suffix, and middle.
As an extension to FIM, IFIM preserves the relative ordering of the original FIM components ($P$ and $S$) and insert the instruction $I$. 
For any given FIM pretraining mode (e.g., PSM), there are three potential IFIM modes (e.g., IPSM, PISM, PSIM).
Let’s use the PISM mode as an example.
Its underlying FIM format (PSM) for input is:
\[
\texttt{<PRE>} \, \circ \, P \, \circ \, \texttt{<SUF>} \,\circ \, S \, \circ \, \texttt{<MID>} 
\]
where $\circ$ denotes concatenation and \texttt{<PRE>}, \texttt{<SUF>}, and \texttt{<MID>} are the special tokens respectively leading the prefix, suffix, and middle sections.
For IFIM, we insert the instruction $I$ at the designated position, i.e, between the prefix $P$ and the \texttt{<SUF>} token for PISM.
Similarly, the instruction $I$ is also paired with a leading special token \texttt{<INS>}.
Therefore, the input of the training sample in PISM mode will be constructed as follows:
\[
\texttt{<PRE>} \, \circ \, P \, \circ \, \texttt{<INS>} \, \circ \, I  \circ \,  \texttt{<SUF>} \circ \, S \, \circ \, \texttt{<MID>}  
\]
The resulted input-output pairs are thus used to train the base code model.
As a new token, \texttt{<INS>} needs to be included in the vocabulary of the base LLM's tokenizer.
However, adding the new tokens to the vocabulary would reinitialize the language model head, risking knowledge forgetting.
To avoid this, we repurpose an existing but rare token from the vocabulary (e.g., the one with the lowest frequency) to serve as the \texttt{<INS>} token.

\subsection{Usage for Code Completion}
An IFIM-trained LLM enables a developer to guide code completion by embedding instructions directly within their code.
Notably, IFIM does not require developers typing instructions for every single completion they want.
Instead, it serves as an effective solution to mitigate the unhelpful code completions caused by the under-specified developer intent.
Since plain text would violate the programming language's syntax, we propose writing instructions as specially marked inline comments.
An example of this is shown in \Cref{fig:usage}.
In this example, a developer using Python could write ``\#!filter out negative numbers'' as the instruction. 
The `\#!' prefix acts as a unique marker that distinguishes an IFIM instruction from a regular comment. 
This allows client-side tools, like IDE plugins, to build user-friendly features around it. 
For example, an IDE could be configured to automatically remove the instructional comment once the generated code is accepted by the user.
When a code completion request is made, the code completion system parses the code around the cursor to extract the prefix and suffix, and identifies the specially marked comment as the instruction.
These three components, prefix, suffix, and instruction, are then formatted into the required IFIM input sequence.
The IFIM-trained LLM receives this input and generates the code to fill the middle, which is then presented to the developer as a completion suggestion.

\section{Experiment Setup}

This section presents the experimental configuration, including training dataset construction, baseline models, evaluation benchmarks, and implementation details.

\subsection{Training Dataset}
IFIM relies on training datasets consisting of code samples with code completion instructions.
Ideally, the IFIM method can work with the production data collected by the real-world code completion applications, where such developer instructions for a specific code completion request can be easily obtained.
Limited by our access to such data, we alternatively create a synthesized one as a proxy to demonstrate its feasibility in our experiments.
It is a well-established strategy that has demonstrated effectiveness in various studies~\cite{wei2023magicoder, luo2023wizardcoder, huang2024opencoder}.
Specifically, we first source various code snippets from OSS-Instruct~\cite{wei2023magicoder} and evol-codealpaca~\cite{luo2023wizardcoder}, two popular open-source instruction datasets.
For each data point, we retain the code samples while removing the original instruction components, yielding 122.9k multi-language samples with 70.0\% written in Python.
For each code snippet, we randomly select 1-3 contiguous lines as the middle span and employ OpenAI GPT-4.1 (gpt-4.1-2025-04-14) as the instruction synthesizer to generate corresponding natural language descriptions for the middle span.
The detailed prompts are available in Appendix~\ref{apx:create}.

Moreover, we perform two data filtering steps on the resulted dataset. To ensure instructions are concise, we decontaminate the dataset by discarding any generated instructions longer than a single sentence.
While some of the remaining instructions may have minor imperfections (e.g., being overly brief or vague), we retain them as they reflect realistic scenarios developers often face. 
Additionally, following~\cite{huang2024opencoder}, we exclude any samples that match problems from common evaluation benchmarks, such as HumanEval~\cite{chen2021evaluating} and MBPP~\cite{austin2021program}.

\subsection{Base Code LLMs}
We evaluate our approach using two popular code LLMs as base models: \textbf{Deepseek-Coder} (6.7B)~\cite{guo2024deepseek} and \textbf{Qwen-2.5-Coder} (7B)~\cite{hui2024qwen25codertechnicalreport}.
This model scale is comparable to those used in production code completion systems (e.g., JetBrains' Mellum~\cite{semenkin2025mellum} at 4B parameters), which prioritize low-latency inference~\cite{wang2023practitioners}.
Each model family includes a base version (pretrained with FIM objectives) and an instruction-tuned version (further fine-tuned on proprietary instruction datasets).
In our experiments, we apply IFIM training to the base versions since they are the LLMs used for code completion.
For a comprehensive comparison, both the original base models and their official instruction-tuned counterparts serve as baselines in our evaluation.

\subsection{Evaluation Benchmarks}
We assess the instruction-following capabilities of our models using benchmarks derived from two complementary ones: IHumanEval from HumanEval-infilling and IRepoMasterEval (IRME) from RepoMasterEval, which respectively represent controlled and real-world evaluation scenarios.
\begin{itemize}[leftmargin=*]
    \item \textbf{HumanEval-infilling}~\cite{fried2022incoder} is a widely-used FIM benchmark constructed by removing middle spans from the canonical solutions in HumanEval.
    We adopt its single-line version, consisting of 1,640 infilling tasks.
    Each task includes a Python function with one random line of missing code, test cases, and docstrings describing the function's functionality.
    While widely adopted, this benchmark has notable limitations for real-world applicability: its function-level scope and well-documented examples differ from practical code completion scenarios. 
    However, considering its popularity, we still adopt it as one evaluation benchmark for our experiments.
    To better simulate real-world conditions and increase the code completion difficulty, we remove the docstrings from all evaluation samples.

    \item \textbf{RepoMasterEval (RME)}~\cite{wu2024repomastereval} provides a more realistic evaluation framework for code completion models, which is used by ByteDance to evaluate their commercial code completion models.
    The pass rate on this benchmark is found to be highly correlated with the online user acceptance rate.
    RepoMasterEval is built upon real-world code repositories, spanning six languages: C++, Go, Java, JavaScript, Python, and TypeScript.
    It extracts the code files from various repositories and processes them into code infilling tasks, i.e., code files with missing middle spans.
    The model outputs are evaluated using the unit tests of the repository, enhanced through mutation testing and manual case additions. Due to computational constraints limiting our models' context windows, we retain the last 20 lines of the prefix and the first 20 lines of the suffix for each sample.
\end{itemize}

Notably, these two benchmarks are not paired with code completion instructions.
To enable the evaluation, we synthesize instructions for each sample using OpenAI GPT-4.1 (prompts available in Appendix~\ref{apx:create}) and then manually revise those unlike real-world instructions.
Given the large sample sizes, we randomly select statistically representative subsets from each benchmark (95\% confidence level, 5\% margin of error, 50\% population proportion), resulting in 312 and 256 samples for HumanEval-infilling and RepoMasterEval, respectively.
The manual revising process involves two authors independently reviewing all generated instructions and revising those that inadequately describe the desired code functionality.
The authors achieve an inter-rater agreement of 0.72 and collaboratively resolve disagreements through discussion.
The final curated instruction benchmarks are respectively named as \textbf{IHumanEval} and \textbf{IRepoMasterEval (IRME)}.
Consent for their release has been obtained from the annotators.

\subsection{Implementation Details}
In our experiments, we use the Huggingface Transformers~\cite{wolf2020transformers} library with PyTorch to implement the models.
The models are trained on a machine with 128 vCPUs, 200GB RAM, and two RTX A6000 GPUs (48GB RAM).
For Deepseek-Coder, we adopt a batch size 2 with 128 accumulation steps and 1216 context length.
For Qwen2.5-Coder, we adopt a batch size 1 with 256 accumulation steps to fit our GPU memory.
These models are trained with an Adafactor optimizer~\cite{shazeer2018adafactor} for 2 epochs.
We set the initial learning rate at 5e-5 with 15 warmup steps and a linear scheduler.
During inference, we use the greedy decoding strategy, and a maximum of 128 tokens are generated.
For FIM mode selection, we adopt the recommended FIM mode, i.e., PSM, based on the default usage example in their official documents.
All training samples are converted to the IFIM format (100\% ratio).
An ablation on alternative data mixing ratios is provided in Appendix~\ref{apx:ratio}.

\begin{table}[t]
  \centering
  \small
\caption{Performance comparison with IFIM and baselines. The setting \emph{w/ ins.} provides each completion request with its corresponding instruction, appended as a comment to the prefix for base and finetuned models, while \emph{w/o ins.} omits the instruction. We abbreviate DeepSeek-Coder as DS-Coder, Qwen2.5-Coder as Qwen-Coder, and IHumanEval as IHE.}
  \label{tab:RQ1}
  \begin{tblr}{
  colsep = 3pt,
  cells = {c},
  cell{2}{1} = {r=12}{},
  cell{2}{2} = {r=6}{},
  cell{5}{3} = {r=3}{},
  cell{8}{2} = {r=6}{},
  cell{11}{3} = {r=3}{},
  cell{14}{1} = {r=12}{},
  cell{14}{2} = {r=6}{},
  cell{17}{3} = {r=3}{},
  cell{20}{2} = {r=6}{},
  cell{23}{3} = {r=3}{},
  vlines,
  hline{1-2,14,26} = {-}{},
  hline{3-5,9-11,15-17,21-23} = {3-6}{},
  hline{6-7,12-13,18-19,24-25} = {4-6}{},
  hline{8,20} = {2-6}{},
}
\textbf{Setting} & \textbf{Model} & \textbf{Version} & \textbf{Mode} & \textbf{IHE} & \textbf{IRME}\\
\emph{w/ ins.} & DS-Coder & base & PSM & 84.6\% & 10.9\%\\
 &  & instruct & prompt & 39.4\% & 2.1\%\\
 &  & finetuned & PSM & 84.6\% & 10.5\%\\
 &  & IFIM & IPSM & \textbf{92.6\%} & \textbf{19.9\%}\\
 &  &  & PISM & \textbf{93.6\%} & \textbf{21.1\%}\\
 &  &  & PSIM & \textbf{92.3\%} & \textbf{19.1\%}\\
 & Qwen-Coder & base & PSM & 91.0\% & 18.4\%\\
 &  & instruct & prompt & 21.8\% & 3.6\%\\
 &  & finetuned & PSM & 90.8\% & 0.4\%\\
 &  & IFIM & IPSM & 78.2\% & 12.9\%\\
 &  &  & PISM & \textbf{95.2\%} & \textbf{19.5\%}\\
 &  &  & PSIM & \textbf{95.8\%} & \textbf{20.3\%}\\
\emph{w/o ins.} & DS-Coder & base & PSM & 68.6\% & 7.4\%\\
 &  & instruct & prompt & 29.8\% & 0.6\%\\
 &  & finetuned & PSM & 72.4\% & 7.0\%\\
 &  & IFIM & IPSM & \textbf{77.9\%} & \textbf{12.9\%}\\
 &  &  & PISM & \textbf{76.6\%} & \textbf{16.0\%}\\
 &  &  & PSIM & \textbf{78.2\%} & \textbf{12.1\%}\\
 & Qwen-Coder & base & PSM & 76.0\% & 10.2\%\\
 &  & instruct & prompt & 20.8\% & 1.9\%\\
 &  & finetuned & PSM & 76.2\% & 0.8\%\\
 &  & IFIM & IPSM & 60.3\% & 9.0\%\\
 &  &  & PISM & \textbf{76.3\%} & \textbf{13.7\%}\\
 &  &  & PSIM & \textbf{76.3\%} & \textbf{13.3\%}\\
\end{tblr}
\end{table}

\section{Results}

\subsection{Feasibility of IFIM}
We assess two aspects to demonstrate the feasibility of IFIM: (1) whether IFIM can improve the instruction-following ability of LLMs, and (2) whether IFIM can maintain the code completion ability when no instruction is given.
We fine-tune base LLMs (DeepSeek-coder and Qwen2.5-coder) using the IFIM training dataset and compare them against three baselines: (1) the original base models (\textbf{base}), (2) the official instruction-tuned versions (\textbf{instruct}), and (3) models fine-tuned on the raw, unprocessed training dataset (\textbf{finetuned}).

In the "with instructions" setting, we pair each completion request with its corresponding instruction. For the base and finetuned models, instructions are written as a comment appended to the end of the prefix, mirroring the real-world practice where developers use inline comments to clarify intent.
For the IFIM variants, instructions are constructed following the IFIM format.
For the instruct variants, we embedded the instruction within a custom prompt designed for the fill-in-the-middle task (available in Appendix~\ref{apx:instruct}).
In the "without instructions" setting, inputs for the base and IFIM models are prefixes and suffixes without modification.
For the instruct variants, however, we use the same prompt that asks the model to perform the fill-in-the-middle task, but without providing the instruction.
The Pass@1 score is computed to assess the quality of the completion.

As reported in~\Cref{tab:RQ1}, the IFIM variants outperform the base models and the finetuned baselines when instructions are provided, across both benchmarks.
In the meantime, when no instructions are provided, with the notable exception of the IPSM mode for Qwen2.5-Coder, the IFIM variants still maintain a comparable, or even better, performance than the base models.
It demonstrates that IFIM not only preserves but can also boost the infilling capability acquired during FIM pre-training.
Moreover, the performance of instruct models aligns with the results in previous studies~\cite{ahmad2025output,gong2024evaluation,pan2024codev}, i.e., existing instruct models impair their capabilities of fill-in-the-middle code completion.

\subsection{Analysis and Ablation Studies}

\begin{table}[t]
\centering
\caption{IHumanEval Pass@1 across model scales (Qwen2.5-Coder family,
PSIM mode for IFIM). \textbf{Bold} indicates the best result within each
setting.}
\label{tab:scale}
\begin{tblr}{
  width = \linewidth,
  colspec = {Q[c,m] Q[c,m] Q[c,m] Q[c,m,1.3] Q[c,m]},
  cells = {c},
  cell{1}{1} = {r=2}{}, 
  cell{1}{2} = {r=2}{}, 
  cell{1}{3} = {c=3}{}, 
  cell{3}{1} = {r=3}{}, 
  cell{6}{1} = {r=3}{}, 
  vlines,
  hline{1,3} = {-}{}, 
  hline{6} = {-}{},   
  hline{9} = {-}{},   
  hline{2} = {3-5}{}, 
  hline{4,5,7,8} = {2-5}{}, 
}
\textbf{Setting} & \textbf{Scale} & \textbf{IHumanEval Pass@1} & & \\
 &  & \textbf{Base} & \textbf{Finetune} & \textbf{IFIM}\\
w/ ins. & 1.5B & 84.0\% & 85.9\% & \textbf{93.9\%}\\
 & 3B & 91.0\% & 87.5\% & \textbf{94.2\%}\\
 & 7B & 91.0\% & 90.8\% & \textbf{95.8\%}\\
w/o ins. & 1.5B & 72.1\% & 73.7\% & \textbf{74.0\%}\\
 & 3B & \textbf{75.0\%} & 74.6\% & 74.7\%\\
 & 7B & 76.0\% & 76.2\% & \textbf{76.3\%}\\
\end{tblr}
\end{table}

\begin{table}[t]
\centering
\caption{Results on the in-the-wild benchmark of 100 real
developer-written instructions mined from GitHub commits. ES denotes
Edit Similarity; the LLM-judge score ranges from 1 to 5. \textbf{Bold}
indicates the best result for each base model.}
\label{tab:wild}
\begin{tblr}{
  width = \linewidth,
  colspec = {Q[c,m,1.3] Q[c,m,1.4] Q[c,m] Q[c,m]},
  cells = {c},
  cell{2}{1} = {r=3}{}, 
  cell{5}{1} = {r=3}{}, 
  vlines,
  hline{1,2} = {-}{}, 
  hline{5} = {-}{},   
  hline{8} = {-}{},   
  hline{3,4,6,7} = {2-4}{}, 
}
\textbf{Base} & \textbf{Setting} & \textbf{ES} & \textbf{LLM-judge}\\
DeepSeek-Coder-6.7B & Baseline (FIM) & 43.7 & 2.43\\
 & Finetuned-comment & 45.3 & 2.33\\
 & IFIM & \textbf{49.5} & \textbf{2.60}\\
Qwen2.5-Coder-7B & Baseline (FIM) & 45.1 & 2.39\\
 & Finetuned-comment & 50.9 & 2.64\\
 & IFIM & \textbf{52.9} & \textbf{2.79}\\
\end{tblr}
\end{table}

\textbf{Performance on Real-developer Instructions}
Both our training data and the instructions in IHumanEval and IRepoMasterEval are synthesized by GPT-4.1. This raises the question of whether the instruction-following capability acquired by IFIM transfers to instructions actually written by developers, whose phrasing, brevity, and precision may differ from an LLM's.
To answer this directly, we construct an evaluation set of genuine developer instructions.
We mine GitHub commits authored after 2025-06, which post-dates the
training cutoff of every model we evaluate, and retain commits in which
a developer deleted a TODO or FIXME comment and implemented code at the
same location within the same commit. Such a comment is an instruction
written by a real developer before the corresponding code existed,
which is precisely the workflow IFIM targets, and the committed code serves as the ground truth.
The construction workflow is introduced in Appendix~\ref{apx:itodoeval}

We report Edit Similarity (ES) against the committed code and an
LLM-judge score for functional consistency, rated from 1 to 5 by
MiMo-V2.5 (the judging protocol and prompt are given in Appendix~\ref{apx:judge}).
Results are shown in Table~\ref{tab:wild}.

IFIM achieves the highest ES and the highest LLM-judge score on both
base models. Notably, the finetuned-comment baseline improves over the
base model on Qwen2.5-Coder but not on DeepSeek-Coder, where its judge
score falls slightly below the base model, whereas IFIM improves on
both. This is consistent with our earlier finding that structural
separation, rather than mere exposure to instruction-bearing data, is
what makes the instruction usable. The result indicates that a model
trained on synthesized instructions generalizes to the distribution of
instructions developers actually write.

\textbf{Generalization across Model Scales.}
Our main experiments use models at the 6.7B--7B scale.
We further train IFIM on Qwen2.5-Coder-1.5B and Qwen2.5-Coder-3B under the identical recipe, with the same set of comparisons.
Table~\ref{tab:scale} reports IHumanEval Pass@1.

Under instructions, IFIM outperforms both the base model and the finetuned baseline at every scale, and the margin does not shrink as the model gets larger. Without instructions, all three variants remain within roughly one point of each other at each scale, confirming that the preservation of FIM capability is not an artifact of a particular model size. We note that the finetuned baseline is unstable across
scales, which improves over the base model at 1.5B but falls behind it
at 3B. In contrast, IFIM is consistent, which again points to the structural separation rather than the additional training as the source of the gain.

\textbf{Impact of IFIM Modes.}
As shown in~\Cref{tab:RQ1}, the instruction-following ability acquired from IFIM training is sensitive to the position of the Instruction component.
Specifically, with the notable exception of the IPSM mode for Qwen2.5-Coder, all other seven IFIM variants outperform their respective base models on both benchmarks when instructions are provided.
Even without instructions, most variants maintain or enhance performance over the base model, confirming the broad effectiveness of IFIM.
Comparing the specific configurations, the PISM and PSIM modes demonstrate comparable performance, with IPSM performs worst.
Based on these observations, we recommend PISM or PSIM as the default configurations for practitioners, since both consistently deliver substantial gains across model families and benchmarks.
Among them, PSIM is preferable when the base model uses a PSM pretraining mode, as it preserves the KV cache for the suffix and thus minimizes inference overhead.
We suggest practitioners avoid placing the instruction before the prefix (IPSM), which underperformed on Qwen2.5-Coder in our experiments.


\begin{table}[t]
    \centering
    \small
    \caption{IFIM vs. Commented FIM: Evaluating the Standalone Instruction Component}
    \label{tab:RQ4}
    \begin{tblr}{
  colsep = 3pt,
  cells = {c},
  cell{2}{1} = {r=3}{},
  cell{5}{1} = {r=3}{},
  vlines,
  hline{1-2,5,8} = {-}{},
  hline{3-4,6-7} = {2-5}{},
}
Model & Version & Mode & IHumanEval & IRME\\
Deepseek-Coder & base & PSM & 84.6\% & 10.9\%\\
 & comment & PSM & 79.0\% & 4.3\%\\
 & IFIM & PISM & \textbf{93.6\%} & \textbf{21.1\%}\\
Qwen2.5-Coder & base & PSM & 91.0\% & 18.4\%\\
 & comment & PSM & 88.5\% & 18.8\%\\
 & IFIM & PSIM & \textbf{95.8\%} & \textbf{20.3\%}
\end{tblr}
\end{table}

\noindent\textbf{Impact of Standalone Instruction Section.}
Compared to treating instructions as a separate, standalone section within the format, a simpler, more straightforward solution would be to train LLMs directly using instructions embedded as in-line comments.
Therefore, we seek to investigate the impact of training on this newly introduced Instruction component, compared with simply learning instruction through inline comments.
Specifically, we derive two training datasets: one processed using the IFIM format and the other by appending the instruction as an inline comment to the prefix.
Using these two derived datasets, we train each base model (DeepSeek-Coder and Qwen2.5-Coder), obtaining the IFIM and comment version.
During inference, the IFIM models are provided with instructed inputs, and the comment variants are provided with commented inputs (i.e., instructions are appended to the prefix as an inline comment).
The results presented in \Cref{tab:RQ4} reveal that simply adding instructions as inline comments is an ineffective and often detrimental strategy.
This confirms a hypothesis: LLMs can be confused when instructional text is mixed with code as comments, negatively impacting their primary code completion capabilities. Therefore, the structural separation of instructions from the code, as implemented in IFIM, is important for the model to properly disambiguate, understand, and benefit from the provided guidance.

\noindent \textbf{Robustness to Noisy Instructions.} Since real-world developer comments are often imperfect, we further evaluate whether IFIM remains effective under noisy instructions.
We inject three types of noise into the IHumanEval instructions: Misspelled (randomly altering one character per instruction), Lengthy (expanding the instruction with two unrelated background sentences), and Ambiguous (randomly removing one word).
Across all three noise types and both model families, IFIM-trained variants consistently outperform their base counterparts—for example, on DeepSeek-Coder under the Misspelled setting, Pass@1 increases from 82.2\% to 91.0\%; on Qwen2.5-Coder under the Lengthy setting, from 85.2\% to 92.2\%.
This indicates that the instruction-following capability acquired through IFIM generalizes beyond clean, well-formed instructions to the kind of noisy directives developers naturally produce.
Full results are reported in Appendix~\ref{apx:robustness}.

\noindent \textbf{Cross-Lingual Generalization.} Although our training data is predominantly Python (70.0\%), the IRepoMasterEval benchmark spans six languages: C++, Go, Java, JavaScript, Python, and TypeScript.
We analyze the per-language results and find that IFIM-trained models consistently outperform the base models on every language where the benchmark provides discriminative signal, i.e., C++, Go, Python, and TypeScript, when instructions are provided.
For instance, on C++, IFIM-trained DeepSeek-Coder achieves 17.9\% Pass@1 compared to the base model's 14.3\%.
This demonstrates that the instruction-following capability induced by IFIM transfers across languages with substantially different syntax, even when those languages are underrepresented in the training data.
The Java and JavaScript slices yield 0\% for both models due to the difficulty of those repo-level tasks and small sample sizes (N=33 and N=7), providing no signal to differentiate methods. 
Per-language results are reported in Appendix~\ref{apx:generalization}.

\noindent \textbf{Practical IDE Considerations.} We further evaluate IFIM along three IDE-relevant dimensions: latency overhead, suggestion diversity, and edit locality. IFIM adds only 0.04\% input-token overhead on IRepoMasterEval, matches the finetuned baseline's suggestion diversity (1.20 vs.\ 1.23 unique-and-correct completions per task), and produces zero cases of unwanted context repetition. Full methodology and results are in Appendix~\ref{apx:practical}.

\begin{table}[t]
    \centering
    \small
    \caption{Evaluating Instruction Alignment on Reversed Instructions for DeepSeek-Coder}
    \label{tab:RQ5}
    \begin{tblr}{
  width = \linewidth,
  colspec = {Q[c,m] Q[c,m,1.5] Q[c,m] Q[c,m]}, 
  cells = {c},
  cell{1}{1} = {r=2}{}, 
  cell{1}{2} = {r=2}{}, 
  cell{1}{3} = {c=2}{}, 
  cell{3}{1} = {r=2}{}, 
  cell{3}{4} = {r=2}{}, 
  cell{5}{1} = {r=2}{}, 
  cell{5}{4} = {r=2}{}, 
  vlines,
  hline{1,3} = {-}{}, 
  hline{5} = {-}{},   
  hline{7} = {-}{},   
  hline{2} = {3-4}{}, 
  hline{4,6} = {2-3}{}, 
}
\textbf{Version} & \textbf{Instruction Type} & \textbf{IHumanEval} & \\
 &  & \textbf{Pass@1} & \textbf{Alignment}\\
base & Standard & 84.6\% & 54.4\%\\
 & Reversed & 59.6\% & \\
ifim & Standard & \textbf{93.6\%} & \textbf{80.9\%}\\
 & Reversed & \textbf{56.7\%} & 
\end{tblr}
\end{table}

\noindent \textbf{Alignment between Instructions and Generated Code.}
Existing benchmarks (IHumanEval, IRME) feature instructions consistent with the code context, making it difficult to attribute IFIM's performance solely to instruction-following capability rather than context understanding.
To decouple these factors, we employed GPT-4.1 to rewrite IHumanEval's instructions, creating deliberate conflicts between the instruction and the original code context.
The prompt and examples are available in~\Cref{fig:examples}.
We evaluated DeepSeek-Coder (base and IFIM) on these conflicting samples using both Pass@1 and human annotation.
Following a similar protocol to our benchmark creation, two authors independently classified each code completion based on whether the code implements the logic from the instruction.
Notably, the annotators are required to only consider the alignment of the code's logic with the instruction's intent, irrespective of the code's functional correctness.
The evaluators achieved an inter-rater agreement of 0.69, and all disagreements were resolved through discussion to reach a final consensus.
As shown in \Cref{tab:RQ5}, the base model frequently ignored the conflicting instructions in favor of the context.
Conversely, IFIM demonstrated superior instruction alignment.
This is evidenced by a sharper drop in Pass@1 on the original tests: by successfully following the modified instructions, IFIM naturally failed the original test cases, confirming its enhanced instruction-following capability.
Moreover, we also performed a case study available in Appendix~\ref{apx:case}, where we find that the IFIM-trained model tends to comply with instructions even though those instructions contradict the original functional context.

\section{Conclusion}
In this paper, we introduced IFIM, a training method that enhances FIM-trained models by extending their training objective to include an explicit instruction component. Our evaluation on popular code LLMs demonstrates that IFIM significantly boosts instruction-following accuracy without compromising the models' original FIM performance.
Moreover, IFIM generalizes robustly to noisy instructions and across multiple programming languages, suggesting broad applicability.
These results validate IFIM as a backward-compatible and effective solution for creating more helpful and responsive code completion systems.

\section*{Limitations}

\noindent \textbf{Scope of Control Baselines.} To isolate the impact of our method, we utilized a controlled 0\% IFIM data ratio to confirm that performance gains stem from instruction integration rather than mere exposure to additional code. However, our current study is scoped to this comparison, focusing on demonstrating the feasibility of this ideas. An important future work is to explore the interplay between IFIM and other alignment techniques. 

\noindent \textbf{Benchmark Scale and Context Truncation.}
Due to the context-window constraints of the evaluated models, IRepoMasterEval samples are truncated to the last 20 lines of the prefix and the first 20 lines of the suffix. Our claims are therefore scoped to the local-context
completion regime. Whether the instruction section retains its
advantage when the model must reconcile an instruction against
thousands of tokens of repository context is an open question, and we leave large-scale long-context evaluation to future work.

\section*{Ethic Consideration}
The proposed method primarily aims to improve developer productivity, which several ethical implications may exist.
As with any generative AI tool, improved code generation capabilities could theoretically be misused to lower the barrier for generating malicious code (e.g., malware or exploits) via natural language instructions.
While our evaluation focused on benign functional correctness, the underlying mechanism of better intent adherence applies broadly.
Future work should investigate safety alignment specifically for instruction-driven code infilling to prevent the generation of harmful content.

\bibliography{references}

@article{ahmad2025output,
  title={From Output to Evaluation: Does Raw Instruction-Tuned Code LLMs Output Suffice for Fill-in-the-Middle Code Generation?},
  author={Ahmad, Wasi Uddin and Majumdar, Somshubra and Ginsburg, Boris},
  journal={arXiv preprint arXiv:2505.18789},
  year={2025}
}

@article{gong2024evaluation,
  title={Evaluation of llms on syntax-aware code fill-in-the-middle tasks},
  author={Gong, Linyuan and Wang, Sida and Elhoushi, Mostafa and Cheung, Alvin},
  journal={arXiv preprint arXiv:2403.04814},
  year={2024}
}

@article{pan2024codev,
  title={Codev-Bench: How Do LLMs Understand Developer-Centric Code Completion?},
  author={Pan, Zhenyu and Cao, Rongyu and Cao, Yongchang and Ma, Yingwei and Li, Binhua and Huang, Fei and Liu, Han and Li, Yongbin},
  journal={arXiv preprint arXiv:2410.01353},
  year={2024}
}

@article{wei2023magicoder,
  title={Magicoder: Empowering code generation with oss-instruct},
  author={Wei, Yuxiang and Wang, Zhe and Liu, Jiawei and Ding, Yifeng and Zhang, Lingming},
  journal={arXiv preprint arXiv:2312.02120},
  year={2023}
}

@article{huang2024opencoder,
  title={Opencoder: The open cookbook for top-tier code large language models},
  author={Huang, Siming and Cheng, Tianhao and Liu, Jason Klein and Hao, Jiaran and Song, Liuyihan and Xu, Yang and Yang, J and Liu, JH and Zhang, Chenchen and Chai, Linzheng and others},
  journal={arXiv preprint arXiv:2411.04905},
  year={2024}
}

@article{luo2023wizardcoder,
  title={Wizardcoder: Empowering code large language models with evol-instruct},
  author={Luo, Ziyang and Xu, Can and Zhao, Pu and Sun, Qingfeng and Geng, Xiubo and Hu, Wenxiang and Tao, Chongyang and Ma, Jing and Lin, Qingwei and Jiang, Daxin},
  journal={arXiv preprint arXiv:2306.08568},
  year={2023}
}

@article{wu2024repomastereval,
  title={Repomastereval: Evaluating code completion via real-world repositories},
  author={Wu, Qinyun and Peng, Chao and Gao, Pengfei and Hu, Ruida and Gan, Haoyu and Jiang, Bo and Tang, Jinhe and Deng, Zhiwen and Guan, Zhanming and Gao, Cuiyun and others},
  journal={arXiv preprint arXiv:2408.03519},
  year={2024}
}

@article{fried2022incoder,
  title={Incoder: A generative model for code infilling and synthesis},
  author={Fried, Daniel and Aghajanyan, Armen and Lin, Jessy and Wang, Sida and Wallace, Eric and Shi, Freda and Zhong, Ruiqi and Yih, Wen-tau and Zettlemoyer, Luke and Lewis, Mike},
  journal={arXiv preprint arXiv:2204.05999},
  year={2022}
}

@article{chen2021evaluating,
  title   = {Evaluating Large Language Models Trained on Code},
  author  = {Mark Chen and Jerry Tworek and Heewoo Jun and Qiming Yuan and Henrique Ponde de Oliveira Pinto and Jared Kaplan and Harri Edwards and Yuri Burda and Nicholas Joseph and Greg Brockman and Alex Ray and Raul Puri and Gretchen Krueger and Michael Petrov and Heidy Khlaaf and Girish Sastry and Pamela Mishkin and Brooke Chan and Scott Gray and Nick Ryder and Mikhail Pavlov and Alethea Power and Lukasz Kaiser and Mohammad Bavarian and Clemens Winter and Philippe Tillet and Felipe Petroski Such and Dave Cummings and Matthias Plappert and Fotios Chantzis and Elizabeth Barnes and Ariel Herbert-Voss and William Hebgen Guss and Alex Nichol and Alex Paino and Nikolas Tezak and Jie Tang and Igor Babuschkin and Suchir Balaji and Shantanu Jain and William Saunders and Christopher Hesse and Andrew N. Carr and Jan Leike and Josh Achiam and Vedant Misra and Evan Morikawa and Alec Radford and Matthew Knight and Miles Brundage and Mira Murati and Katie Mayer and Peter Welinder and Bob McGrew and Dario Amodei and Sam McCandlish and Ilya Sutskever and Wojciech Zaremba},
  year    = {2021},
  journal = {arXiv preprint arXiv: 2107.03374}
}

@inproceedings{shazeer2018adafactor,
  title={Adafactor: Adaptive learning rates with sublinear memory cost},
  author={Shazeer, Noam and Stern, Mitchell},
  booktitle={International Conference on Machine Learning},
  pages={4596--4604},
  year={2018},
  organization={PMLR}
}

@article{sun2025don,
  title={Don’t complete it! Preventing unhelpful code completion for productive and sustainable neural code completion systems},
  author={Sun, Zhensu and Du, Xiaoning and Song, Fu and Wang, Shangwen and Ni, Mingze and Li, Li and Lo, David},
  journal={ACM Transactions on Software Engineering and Methodology},
  volume={34},
  number={1},
  pages={1--22},
  year={2025},
  publisher={ACM New York, NY}
}

@inproceedings{mozannar2024reading,
  title={Reading between the lines: Modeling user behavior and costs in AI-assisted programming},
  author={Mozannar, Hussein and Bansal, Gagan and Fourney, Adam and Horvitz, Eric},
  booktitle={Proceedings of the 2024 CHI Conference on Human Factors in Computing Systems},
  pages={1--16},
  year={2024}
}

@article{guo2025deepseek,
  title={Deepseek-r1: Incentivizing reasoning capability in llms via reinforcement learning},
  author={Guo, Daya and Yang, Dejian and Zhang, Haowei and Song, Junxiao and Zhang, Ruoyu and Xu, Runxin and Zhu, Qihao and Ma, Shirong and Wang, Peiyi and Bi, Xiao and others},
  journal={arXiv preprint arXiv:2501.12948},
  year={2025}
}

@article{yang2025qwen3,
  title={Qwen3 technical report},
  author={Yang, An and Li, Anfeng and Yang, Baosong and Zhang, Beichen and Hui, Binyuan and Zheng, Bo and Yu, Bowen and Gao, Chang and Huang, Chengen and Lv, Chenxu and others},
  journal={arXiv preprint arXiv:2505.09388},
  year={2025}
}

@inproceedings{zheng2023codegeex,
  title={Codegeex: A pre-trained model for code generation with multilingual benchmarking on humaneval-x},
  author={Zheng, Qinkai and Xia, Xiao and Zou, Xu and Dong, Yuxiao and Wang, Shan and Xue, Yufei and Shen, Lei and Wang, Zihan and Wang, Andi and Li, Yang and others},
  booktitle={Proceedings of the 29th ACM SIGKDD Conference on Knowledge Discovery and Data Mining},
  pages={5673--5684},
  year={2023}
}

@article{nijkamp2022codegen,
  title={Codegen: An open large language model for code with multi-turn program synthesis},
  author={Nijkamp, Erik and Pang, Bo and Hayashi, Hiroaki and Tu, Lifu and Wang, Huan and Zhou, Yingbo and Savarese, Silvio and Xiong, Caiming},
  journal={arXiv preprint arXiv:2203.13474},
  year={2022}
}

@article{deepseek-ai2024deepseek0coder0v20,
  title   = {DeepSeek-Coder-V2: Breaking the Barrier of Closed-Source Models in Code Intelligence},
  author  = {DeepSeek-AI and Qihao Zhu and Daya Guo and Zhihong Shao and Dejian Yang and Peiyi Wang and Runxin Xu and Y. Wu and Yukun Li and Huazuo Gao and Shirong Ma and Wangding Zeng and Xiao Bi and Zihui Gu and Hanwei Xu and Damai Dai and Kai Dong and Liyue Zhang and Yishi Piao and Zhibin Gou and Zhenda Xie and Zhewen Hao and Bingxuan Wang and Junxiao Song and Deli Chen and Xin Xie and Kang Guan and Yuxiang You and Aixin Liu and Qiushi Du and Wenjun Gao and Xuan Lu and Qinyu Chen and Yaohui Wang and Chengqi Deng and Jiashi Li and Chenggang Zhao and Chong Ruan and Fuli Luo and Wenfeng Liang},
  year    = {2024},
  journal = {arXiv preprint arXiv: 2406.11931}
}

@article{team2024codegemma,
  title={Codegemma: Open code models based on gemma},
  author={Team, CodeGemma and Zhao, Heri and Hui, Jeffrey and Howland, Joshua and Nguyen, Nam and Zuo, Siqi and Hu, Andrea and Choquette-Choo, Christopher A and Shen, Jingyue and Kelley, Joe and others},
  journal={arXiv preprint arXiv:2406.11409},
  year={2024}
}

@inproceedings{sagtani2025improving,
  title={Improving fim code completions via context \& curriculum based learning},
  author={Sagtani, Hitesh and Mehrotra, Rishabh and Liu, Beyang},
  booktitle={Proceedings of the Eighteenth ACM International Conference on Web Search and Data Mining},
  pages={801--810},
  year={2025}
}

@article{gong2025structure,
  title={Structure-Aware Fill-in-the-Middle Pretraining for Code},
  author={Gong, Linyuan and Cheung, Alvin and Elhoushi, Mostafa and Wang, Sida},
  journal={arXiv preprint arXiv:2506.00204},
  year={2025}
}

@article{ding2024horizon,
  title={Horizon-length prediction: Advancing fill-in-the-middle capabilities for code generation with lookahead planning},
  author={Ding, Yifeng and Ding, Hantian and Wang, Shiqi and Sun, Qing and Kumar, Varun and Wang, Zijian},
  journal={arXiv preprint arXiv:2410.03103},
  year={2024}
}

@article{bavarian2022efficient,
  title={Efficient training of language models to fill in the middle},
  author={Bavarian, Mohammad and Jun, Heewoo and Tezak, Nikolas and Schulman, John and McLeavey, Christine and Tworek, Jerry and Chen, Mark},
  journal={arXiv preprint arXiv:2207.14255},
  year={2022}
}

@article{wei2021finetuned,
  title={Finetuned language models are zero-shot learners},
  author={Wei, Jason and Bosma, Maarten and Zhao, Vincent Y and Guu, Kelvin and Yu, Adams Wei and Lester, Brian and Du, Nan and Dai, Andrew M and Le, Quoc V},
  journal={arXiv preprint arXiv:2109.01652},
  year={2021}
}

@misc{hui2024qwen25codertechnicalreport,
      title={Qwen2.5-Coder Technical Report}, 
      author={Binyuan Hui and Jian Yang and Zeyu Cui and Jiaxi Yang and Dayiheng Liu and Lei Zhang and Tianyu Liu and Jiajun Zhang and Bowen Yu and Keming Lu and Kai Dang and Yang Fan and Yichang Zhang and An Yang and Rui Men and Fei Huang and Bo Zheng and Yibo Miao and Shanghaoran Quan and Yunlong Feng and Xingzhang Ren and Xuancheng Ren and Jingren Zhou and Junyang Lin},
      year={2024},
      eprint={2409.12186},
      archivePrefix={arXiv},
      primaryClass={cs.CL},
      url={https://arxiv.org/abs/2409.12186}, 
}

@article{guo2024deepseek,
  title={DeepSeek-Coder: When the Large Language Model Meets Programming--The Rise of Code Intelligence},
  author={Guo, Daya and Zhu, Qihao and Yang, Dejian and Xie, Zhenda and Dong, Kai and Zhang, Wentao and Chen, Guanting and Bi, Xiao and Wu, Yu and Li, YK and others},
  journal={arXiv preprint arXiv:2401.14196},
  year={2024}
}

@article{austin2021program,
  title   = {Program Synthesis with Large Language Models},
  author  = {Jacob Austin and Augustus Odena and Maxwell Nye and Maarten Bosma and Henryk Michalewski and David Dohan and Ellen Jiang and Carrie Cai and Michael Terry and Quoc Le and Charles Sutton},
  year    = {2021},
  journal = {arXiv preprint arXiv: 2108.07732}
}

@inproceedings{wolf2020transformers,
  title={Transformers: State-of-the-art natural language processing},
  author={Wolf, Thomas and Debut, Lysandre and Sanh, Victor and Chaumond, Julien and Delangue, Clement and Moi, Anthony and Cistac, Pierric and Rault, Tim and Louf, Remi and Funtowicz, Morgan and others},
  booktitle={Proceedings of the 2020 conference on empirical methods in natural language processing: system demonstrations},
  pages={38--45},
  year={2020}
}

@inproceedings{wang2023practitioners,
  title={How practitioners expect code completion?},
  author={Wang, Chaozheng and Hu, Junhao and Gao, Cuiyun and Jin, Yu and Xie, Tao and Huang, Hailiang and Lei, Zhenyu and Deng, Yuetang},
  booktitle={Proceedings of the 31st ACM Joint European Software Engineering Conference and Symposium on the Foundations of Software Engineering},
  pages={1294--1306},
  year={2023}
}

@misc{semenkin2025mellum,
  author       = {Semenkin, Anton},
  title        = {Mellum: How We Trained a Model to Excel in Code Completion},
  howpublished = {JetBrains AI Blog},
  year         = {2025},
  month        = apr,
  url          = {https://blog.jetbrains.com/ai/2025/04/mellum-how-we-trained-a-model-to-excel-in-code-completion/},
  note         = {Accessed: 2026-05-12}
}

\newpage
\appendix

\begin{figure*}[t]
    \centering
    \caption{A visual comparison of outputs produced by base and IFIM DeepSeek-Coder given conflicting instructions. The checkmark (\cmark) indicates that the generated code aligns with the intent of the Conflicting Instruction, while the cross mark (\xmark) indicates that it does not.}
    \label{fig:examples}
    \vspace{2mm}
    
    \begin{minipage}{0.9\textwidth}
    \begin{tcolorbox}[
        title={\textbf{Example 1:} HumanEval/143/L3},
        colback=black!5, colframe=black!75, fonttitle=\small\ttfamily,
        boxsep=0pt, left=4pt, right=4pt, top=2pt, bottom=4pt
    ]
    \small
    {\color{mygray}\textbf{Original Instruction:}} Check if the length of the word is 1.\\
    {\color{myblue}\textbf{Conflicting Instruction:}} Check if the word starts with 'a'.
    \tcblower
    \textbf{Base model:} \verb|if word[0] != 'a':| \xmark
    
    \textbf{IFIM model:} \verb|if word[0] == 'a':| \cmark
    \end{tcolorbox}
    \end{minipage}
    
    \vspace{3mm}
    
    \begin{minipage}{0.9\textwidth}
    \begin{tcolorbox}[
        title={\textbf{Example 2:} HumanEval/140/L12},
        colback=black!5, colframe=black!75, fonttitle=\small\ttfamily,
        boxsep=0pt, left=4pt, right=4pt, top=2pt, bottom=4pt
    ]
    \small
    {\color{mygray}\textbf{Original Instruction:}} Append the current non-space character.\\
    {\color{myblue}\textbf{Conflicting Instruction:}} Append an underscore character.
    \tcblower
    \textbf{Base model:} \verb|new_text += text[i]| \xmark
    
    \textbf{IFIM model:} \verb|new_text += "_"| \cmark
    \end{tcolorbox}
    \end{minipage}

    \vspace{3mm}

    \begin{minipage}{0.9\textwidth}
    \begin{tcolorbox}[
        title={\textbf{Example 3:} HumanEval/3/L4},
        colback=black!5, colframe=black!75, fonttitle=\small\ttfamily,
        boxsep=0pt, left=4pt, right=4pt, top=2pt, bottom=4pt
    ]
    \small
    {\color{mygray}\textbf{Original Instruction:}} Check if the balance is less than zero.\\
    {\color{myblue}\textbf{Conflicting Instruction:}} Check if the balance is greater than zero.
    \tcblower
    \textbf{Base model:} \verb|if balance < 0:| \xmark
    
    \textbf{IFIM model:} \verb|if balance > 0:| \cmark
    \end{tcolorbox}
    \end{minipage}

    \vspace{3mm}

    \begin{minipage}{0.9\textwidth}
    \begin{tcolorbox}[
        title={\textbf{Example 4:} HumanEval/13/L0},
        colback=black!5, colframe=black!75, fonttitle=\small\ttfamily,
        boxsep=0pt, left=4pt, right=4pt, top=2pt, bottom=4pt
    ]
    \small
    {\color{mygray}\textbf{Original Instruction:}} Repeat the loop as long as b is not zero.\\
    {\color{myblue}\textbf{Conflicting Instruction:}} Repeat the loop as long as a is not zero.
    \tcblower
    \textbf{Base model:} \verb|while b != 0:| \xmark
    
    \textbf{IFIM model:} \verb|while a != 0:| \cmark
    \end{tcolorbox}
    \end{minipage}

    \vspace{3mm}

    \begin{minipage}{0.9\textwidth}
    \begin{tcolorbox}[
        title={\textbf{Example 5:} HumanEval/63/L4},
        colback=black!5, colframe=black!75, fonttitle=\small\ttfamily,
        boxsep=0pt, left=4pt, right=4pt, top=2pt, bottom=4pt
    ]
    \small
    {\color{mygray}\textbf{Original Instruction:}} Check if n equals to 2.\\
    {\color{myblue}\textbf{Conflicting Instruction:}} Check if n equals to 3.
    \tcblower
    \textbf{Base model:} \verb|if n == 2:| \xmark
    
    \textbf{IFIM model:} \verb|if n == 3:| \cmark
    \end{tcolorbox}
    \end{minipage}
\end{figure*}

\section{Prompt for annotating the training and evaluation datasets}\label{apx:create}
We present the entire code sample to the LLM, but wrap the middle part with <explain></explain> tags and request a concise, one-sentence functional description.
The prompt is structured as follows:
\begin{tcolorbox}[size=title]
    {\textbf{System Prompt:}}
    
    You are a senior software engineer. When given a code snippet containing sections marked with <explain></explain> tags, write a single, concise instruction that explicitly describes the functional purpose of the code to be implemented within the tagged area. Focus on what needs to be achieved (e.g., inputs, outputs, logic) without prescribing how to implement it (e.g., specific methods, libraries). Ensure clarity and brevity so a developer can directly translate the instruction into code. Only write the instruction, no other text.\\

    {\textbf{User Prompt:}}

    Explain the code in the <explain></explain> tags using one simple sentence: \verb|```|\{language\}
    
    \{code\}
    
    \verb|```|
\end{tcolorbox}

\section{Prompt for the instruct models in our experiments}\label{apx:instruct}
The models are provided with the following prompt:

\begin{tcolorbox}[size=title]
    {\textbf{Prompt:}}
    
    Below is a incomplete code implementation and the description of its missing part.
    
* Incomplete Code:

\verb|```|\{language\}

\{code\}

\verb|```|

* Description:
\{instruction\}

Please write code to fill the [MASK] (multiple lines of code may be masked out).
Your code should only include the code to fill the [MASK].
Do not repeat the context of the [MASK].
\end{tcolorbox}

\section{Prompt for the LLM Judge}\label{apx:judge}

Since mined samples carry no test cases, we complement Exact Match and Edit Similarity with an LLM judge that rates functional consistency between the generated completion and the instruction. We use MiMo-V2.5 with temperature 0. For each sample, the judge receives the instruction, the surrounding context, the developer-written reference, and the model completion, and returns a score from 1 to 5. The prompt is structured as follows:

\begin{tcolorbox}[size=title]
    {\textbf{Prompt:}}

    You are evaluating code completions. A developer left this instruction (originally a TODO comment) at a location in a file, and a model filled in the missing code.\\

    INSTRUCTION: \{instruction\}\\

    CODE BEFORE the missing part (tail):

    \verb|```|

    \{prefix\}

    \verb|```|

    CODE AFTER the missing part (head):

    \verb|```|

    \{suffix\}

    \verb|```|

    REFERENCE (what the developer actually wrote):

    \verb|```|

    \{reference\}

    \verb|```|

    MODEL COMPLETION:

    \verb|```|

    \{generation\}

    \verb|```|\\

    Score the MODEL COMPLETION from 1 to 5:

    5 = functionally equivalent to the reference, fulfills the instruction;

    4 = fulfills the instruction with a valid alternative approach, fits the context (would plausibly work);

    3 = partially fulfills the instruction, or right direction with clear flaws;

    2 = related to the instruction but mostly wrong or does not fit the context;

    1 = unrelated, empty, degenerate, or contradicts the instruction.\\

    Answer with ONLY a JSON object: \{"score": <1-5>\}
\end{tcolorbox}

\section{Impact of IFIM Training Data Ratio}
\label{apx:ratio}

\begin{figure}[t]
  \centering
  \includegraphics[width=\linewidth]{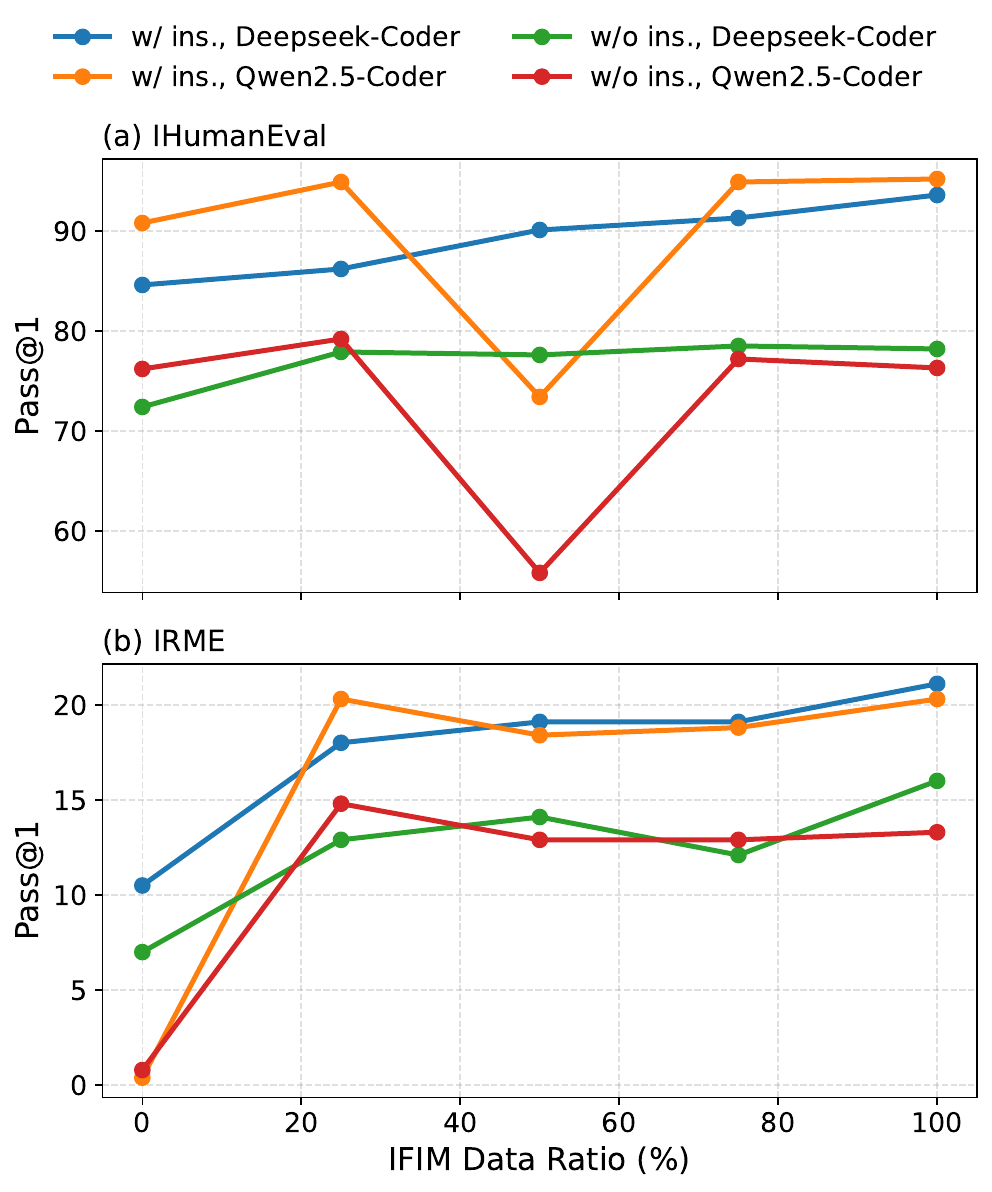}
  \caption{Impact of IFIM Training Data Ratio in Training on Model Performance}
  \label{fig:ratio}
\end{figure}

A natural design question for IFIM is how much of the training data should be converted to the IFIM format, as opposed to retained in the standard FIM format native to the base model. To investigate this, we experimented with five different data mixtures, converting 0\%, 25\%, 50\%, 75\%, and 100\% of the training samples into the IFIM format, with the remainder kept in the standard FIM format. For this experiment, we used the best-performing IFIM mode for each model: PISM for DeepSeek-Coder and PSIM for Qwen2.5-Coder. The results are shown in~\Cref{fig:ratio}.

Overall, a full conversion of the training data to the IFIM format (100\%) proves to be an effective strategy across both model families and benchmarks.
However, the relationship between data ratio and model performance is not strictly linear and reveals complex trade-offs. The most striking example is Qwen2.5-Coder, where the 50\% ratio causes a notable performance drop on IHumanEval (73.4\%) compared to both lower and higher ratios, suggesting that the optimal mixture is model-dependent.
We therefore recommend practitioners default to a 100\% IFIM conversion, which consistently delivers strong results without requiring model-specific tuning of the data mixture.

\section{Case Study for Alignment}\label{apx:case}
We further qualitatively demonstrate the instruction-following behaviors between base and IFIM models using five real examples, shown in~\Cref{fig:examples}.
In each case, the base model tends to ignore the commented instruction, defaulting to the logic suggested by the original function context.
For instance, in Example 5 (HumanEval/63/L4), given the conflicting instruction that requires to check if n equals to 3, the base model still generates `if n == 2:', the code expected by the code context.
Conversely, the IFIM model adheres to the new instruction, producing `if n == 3:'.
These examples provide qualitative evidence that IFIM training genuinely enhances the model's ability to prioritize and follow explicit natural language instructions, even when they contradict the surrounding code context.

\section{Robustness to Noisy Instructions}
\label{apx:robustness}

Real-world developer comments are often imperfect: they may contain 
typos, be unnecessarily verbose, or omit relevant details.
To assess whether IFIM remains effective under such conditions, we construct three noisy variants of the IHumanEval instructions:

\begin{itemize}
    \item \textbf{Misspelled}: randomly alter one character in each 
    instruction, simulating typographical errors.
    \item \textbf{Lengthy}: expand each instruction with two unrelated background sentences, simulating verbose comments that mix intent with extraneous context.
    \item \textbf{Ambiguous}: randomly remove one word from each 
    instruction, simulating the kind of elision common in informal 
    developer notes.
\end{itemize}

Both DeepSeek-Coder and Qwen2.5-Coder are evaluated in their base and IFIM-trained variants. The IFIM variants use the best-performing 
mode for each model family (PISM for DeepSeek-Coder, PSIM for 
Qwen2.5-Coder). Results are shown in Table~\ref{tab:robustness}.

\begin{table}[t]
  \centering
  \small
  \caption{Pass@1 on IHumanEval under three noise injection settings. 
  \textbf{Bold} indicates IFIM outperforms its base counterpart.}
  \label{tab:robustness}
  \begin{tblr}{
  colsep = 3pt,
  cells = {c},
  cell{2}{1} = {r=6}{},
  cell{2}{2} = {r=2}{},
  cell{4}{2} = {r=2}{},
  cell{6}{2} = {r=2}{},
  cell{8}{1} = {r=6}{},
  cell{8}{2} = {r=2}{},
  cell{10}{2} = {r=2}{},
  cell{12}{2} = {r=2}{},
  vlines,
  hline{1-2,8,14} = {-}{},
  hline{3,5,7,9,11,13} = {3-4}{},
  hline{4,6,10,12} = {2-4}{},
}
\textbf{Model} & \textbf{Setting} & \textbf{Version} & \textbf{Pass@1}\\
DeepSeek-Coder & Misspelled & base & 82.2\%\\
 &  & IFIM & \textbf{91.0\%}\\
 & Lengthy & base & 82.9\%\\
 &  & IFIM & \textbf{91.6\%}\\
 & Ambiguous & base & 77.4\%\\
 &  & IFIM & \textbf{84.5\%}\\
Qwen2.5-Coder & Misspelled & base & 89.9\%\\
 &  & IFIM & \textbf{93.3\%}\\
 & Lengthy & base & 85.2\%\\
 &  & IFIM & \textbf{92.2\%}\\
 & Ambiguous & base & 80.0\%\\
 &  & IFIM & \textbf{86.3\%}\\
\end{tblr}
\end{table}

Across all three noise types and both model families, IFIM-trained 
variants consistently outperform their base counterparts.
The improvement is largest under the Lengthy setting (e.g., +8.7 points on DeepSeek-Coder, +7.0 points on Qwen2.5-Coder), suggesting that  the structural separation of instructions in IFIM helps the model 
isolate intent from surrounding noise. Even under the most degrading 
Ambiguous setting, where one word is randomly removed, IFIM retains 
a consistent advantage of 6--7 points over the base models. These 
results indicate that the instruction-following capability acquired 
through IFIM generalizes beyond clean, well-formed instructions to 
the kind of imperfect directives developers naturally produce in 
practice.

\section{Per-Language Results on IRepoMasterEval}
\label{apx:generalization}

IRepoMasterEval spans six languages: C++, Go, Java, JavaScript, Python, 
and TypeScript.
Table~\ref{tab:cross-lingual} reports per-language 
Pass@1 for DeepSeek-Coder (base vs. IFIM-trained, PISM mode) under both ``with instructions'' and ``without instructions'' settings.

\begin{table*}[t]
  \centering
  \small
  \caption{Per-language Pass@1 on IRepoMasterEval (DeepSeek-Coder, PISM mode). \textbf{Bold} indicates IFIM outperforms its base 
  counterpart on that language.}
  \label{tab:cross-lingual}
  \begin{tblr}{
  colsep = 3pt,
  cells = {c},
  cell{2}{1} = {r=12}{},
  cell{2}{2} = {r=6}{},
  cell{8}{2} = {r=6}{},
  cell{14}{1} = {r=12}{},
  cell{14}{2} = {r=6}{},
  cell{20}{2} = {r=6}{},
  vlines,
  hline{1-2,14,26} = {-}{},
  hline{3-7,9-13,15-19,21-25} = {3-5}{},
  hline{8,20} = {2-5}{},
}
\textbf{Setting} & \textbf{Version} & \textbf{Language} & \textbf{N} & \textbf{Pass@1}\\
Inputs with instructions & base & C++ & 28 & 14.3\%\\
 &  & Go & 109 & 5.5\%\\
 &  & Java & 33 & 0.0\%\\
 &  & JavaScript & 7 & 0.0\%\\
 &  & Python & 38 & 34.2\%\\
 &  & TypeScript & 50 & 10.0\%\\
 & IFIM & C++ & 28 & \textbf{17.9\%}\\
 &  & Go & 109 & \textbf{12.8\%}\\
 &  & Java & 33 & 0.0\%\\
 &  & JavaScript & 7 & 0.0\%\\
 &  & Python & 38 & \textbf{50.0\%}\\
 &  & TypeScript & 50 & \textbf{22.0\%}\\
Inputs without instructions & base & C++ & 28 & 10.7\%\\
 &  & Go & 109 & 3.7\%\\
 &  & Java & 33 & 0.0\%\\
 &  & JavaScript & 7 & 0.0\%\\
 &  & Python & 38 & \textbf{23.7\%}\\
 &  & TypeScript & 50 & 6.0\%\\
 & IFIM & C++ & 28 & \textbf{14.3\%}\\
 &  & Go & 109 & \textbf{11.9\%}\\
 &  & Java & 33 & 0.0\%\\
 &  & JavaScript & 7 & 0.0\%\\
 &  & Python & 38 & 21.1\%\\
 &  & TypeScript & 50 & \textbf{20.0\%}\\
\end{tblr}
\end{table*}

On the four languages with discriminative signal (C++, Go, Python, TypeScript), IFIM consistently outperforms the base model when instructions are provided. The instruction-following capability transfers across all four languages despite the training data being dominated by Python (70.0\%).
When no instructions are provided, IFIM maintains or improves performance on three of the four languages, with Python showing a modest decrease (from 23.7\% to 21.1\%)  that is offset by substantial gains in the instruction-conditioned setting.

\section{Practical IDE Considerations}
\label{apx:practical}

Beyond Pass@1, deploying IFIM in a real IDE raises three practical 
concerns: latency overhead, suggestion diversity, and edit locality. 
We report the methodology and detailed results for each below.

\paragraph{Latency Overhead.}
A code completion model must respond within tight latency budgets 
to remain usable.
IFIM introduces exactly one additional special 
token (\texttt{<INS>}) per request beyond the natural-language tokens of the instruction itself.
To quantify the structural overhead introduced by IFIM (independent of instruction length, which is under the developer's control), we measure the proportion of total input tokens contributed by this single special token across all IRepoMasterEval samples.
The overhead averages 0.04\% of the total input tokens, making the structural cost of IFIM negligible relative to typical code-completion latency budgets.
The instruction text itself contributes additional tokens proportional to the instruction length, but this is identical to the cost of writing the comment in the base setting.

\paragraph{Suggestion Diversity.}
Code completion systems often present multiple candidate completions 
to the developer. A common concern with fine-tuned models is that 
they may collapse to a narrow output distribution, reducing the 
utility of multi-suggestion interfaces. To assess whether IFIM 
preserves output diversity, we sample five completions per task on 
IHumanEval using DeepSeek-Coder with temperature $t = 0.2$ and count 
the number of \textit{unique-and-correct} suggestions per task 
(i.e., suggestions that pass the unit tests and are textually 
distinct from one another). IFIM produces an average of 1.20 such 
suggestions per task, closely matching the finetuned baseline's 
1.23. This indicates that IFIM does not meaningfully reduce 
suggestion diversity compared to standard fine-tuning.

\paragraph{Edit Locality.}
A common failure mode of standard instruction-tuned models in 
code completion settings is re-emitting tokens already present in 
the surrounding context.
For example, repeating the leading tokens of the suffix or trailing tokens of the prefix within the generated middle segment.
Such redundancy breaks edit locality and forces the IDE to perform post-hoc deduplication.
We check for this failure mode programmatically by scanning each IFIM-generated completion for overlap with the leading tokens of the provided suffix or the trailing tokens of the prefix.
Across both IHumanEval and IRepoMasterEval, our script initially flagged 4.2\% of IFIM-generated samples as containing potential overlap.
Subsequent manual inspection of all flagged cases confirmed \textbf{zero} actual instances of unwanted context repetition.
This indicates that IFIM avoids the boundary-blurring behaviors common to standard instruction-tuned variants, preserving the clean edit locality expected of FIM-trained models.

\section{Marker Parsing in IDEs}\label{apx:marker}
IFIM relies on a designated marker (e.g., \#! in Python) to distinguish instruction comments from ordinary comments. The parsing logic is minimal: it requires only that the IDE's lexer recognize comments and check for the predefined marker prefix, a capability supported by all modern IDEs. The marker is configurable per language and can be made more complex (e.g., \#!ai:) to reduce the risk of accidental matches with naturally occurring comments in the wild.

\section{Construction of ITodoEval}\label{apx:itodoeval}

To evaluate IFIM on instructions genuinely written by developers, we construct ITodoEval, a benchmark mined from GitHub commits in which a developer deleted a \texttt{TODO}/\texttt{FIXME} comment and implemented the corresponding code at the same location in the same commit. The deleted comment is a real instruction written \emph{before} the code existed---precisely the workflow IFIM targets---and the committed code serves as the ground truth. Each sample is a quadruple (\textit{instruction}, \textit{prefix}, \textit{middle}, \textit{suffix}): the instruction is the TODO text with its marker prefix (e.g., \texttt{\# TODO:}) stripped and the original wording preserved; the middle is the code added in the commit; the prefix and suffix are up to 20 lines of context on each side, taken from the post-commit version of the file. The instruction is delivered exclusively through the IFIM channel: the mined context is guaranteed not to contain the deleted TODO line, and the middle is verified not to appear verbatim in the context.

\paragraph{Repository and commit mining.}
We query the GitHub Search API for repositories with at least 100 stars, a permissive license (MIT, Apache-2.0, or BSD-2/3-Clause), and push activity within the last year, across ten programming languages, yielding a pool of 1,770 repositories. Each repository is shallow-cloned and its commits are scanned. To prevent training-data leakage, we only keep commits authored after June 2025, post-dating the training cutoff of every evaluated model; because \texttt{git} filters history by commit date while rebased commits may carry older author dates, we additionally re-validate the author date of every sample. Merge commits and commits from automated accounts (e.g., \texttt{dependabot}, \texttt{renovate}) are excluded. In total, 622 repositories and 463,910 commits were scanned.

\paragraph{Candidate extraction.}
For every modified source file in a commit, we locate diff hunks in which (i) a deleted line matches a \texttt{TODO}/\texttt{FIXME} line comment (\texttt{\#} or \texttt{//} style) and (ii) the same hunk adds a contiguous block of 1--15 lines containing executable code (not exclusively comments or docstrings). The contiguous added block closest to the deleted comment becomes the middle. Multi-line TODO comments are captured by absorbing contiguous deleted comment lines that follow the marker line. For Python files, the complete post-commit file must parse with \texttt{ast.parse}.

\paragraph{Filtering and deduplication.}
Candidates then pass through a cascade of filters, each recording its rejection reason: the instruction must contain at least five words of English text and express an actionable intent (pure cross-references such as ``see issue \#123'', bare URLs, and attribution notes are rejected by pattern heuristics that require an action cue); the combined context must contain at least five non-empty lines. We deduplicate instructions exactly and after normalization, and cap the yield at three samples per commit, five per file, and 25 per repository so that no single project dominates. This stage reduced 4,825 raw candidates to 1,803.

\paragraph{Curation and verification.}
From the 1,803 candidates we curated the final benchmark in three stages. First, a heuristic pre-ranking (favoring multi-line targets and concise instructions, with per-language quotas) selected a pool of 435 candidates. Second, an LLM screener (Claude) scored each candidate for instruction--code consistency, and the top-scoring candidates were selected under language-diversity and per-repository constraints. Third, every selected sample underwent an adversarial verification pass against five criteria: the instruction is an actionable directive; it is understandable at the code location without referring to code deleted by the commit; the middle genuinely implements the instruction; the task is completable from the given context; and the instruction does not spell out the middle verbatim. Samples failing any criterion were replaced by verified alternates, and the resulting set was manually reviewed by two authors. The final ITodoEval benchmark contains 100 samples from 75 repositories.

\end{document}